\documentclass[12pt,prd,tightenlines,floatfix,preprintnumbers,nofootinbib,eqsecnum]{revtex4}
 \usepackage[dvips,final]{graphicx}
  \usepackage{bm}
   \usepackage{amsmath}
    \usepackage{amssymb}
     \usepackage{pifont}
      \usepackage{epsfig}
       \usepackage{wrapfig}

\newcommand{\va}[1]{\langle{#1}\rangle}                   
\newcommand{\gev}[1]{\relax\ifmmode{\text{GeV}^{#1}}      
                     \else{GeV$^{#1}${ }}\fi}             
\newcommand{\Gev}{\relax\ifmmode{\text{GeV}}              
                     \else{GeV{ }}\fi}                    
\newcommand{\Mev}{\relax\ifmmode{\text{MeV}}              
                     \else{MeV{ }}\fi}                    
\def\as{\relax\ifmmode \alpha_s\else{$ \alpha_s${ }}\fi}  

\begin{document}
\preprint{\hbox{RUB-TPII-03/19}}
\title{Mapping out the quark structure of hadrons in QCD\footnote{%
       Invited contribution to the Festschrift
       in honor of Prof.~Anatoly Efremov's 70th birthday}}

\author{Alexander~P.~Bakulev}%
 \email{bakulev@thsun1.jinr.ru}
\author{S.~V.~Mikhailov}%
 \email{mikhs@thsun1.jinr.ru}
\author{N.~G.~Stefanis\footnote{On leave of absence from
 Institut f\"ur Theoretische Physik II, Ruhr-Universit\"at Bochum,
 D-44780 Bochum, Germany}}
 \email{stefanis@tp2.ruhr-uni-bochum.de}
\affiliation{%
 Bogoliubov Laboratory of Theoretical Physics,
 JINR, 141980, Moscow Region, Dubna, Russia}%
\vspace {10mm}

\begin{abstract}
In the context of QCD sum rules with non-local condensates we present a
pion distribution amplitude, which is double-humped with its end-points
$x \to (0\, ,1)$ strongly suppressed, and show that it matches the
CLEO experimental data on the pion-photon transition at the 1$\sigma$
level accuracy, being also in compliance with the CELLO data.
We also include some comments on the nucleon distribution amplitude
and the nucleon evolution equation.
\end{abstract}

\maketitle

\section{A tribute to Prof.~Efremov's celebration of his
         70\lowercase{th} birthday}

Prof.~Efremov gives us the opportunity to point out in this Festschrift
the influence of his work on our own research activities.

A.~V.~Efremov is one of the inventors of factorization theorems in
quantum field theory that are particularly indispensable in applying
perturbative QCD in inclusive \cite{ER80tmf1} and exclusive
reactions \cite{ER80,ER80tmf} involving hadrons.
Without these tools, the experimental verification of QCD would
constitute an intractable task.
Together with his then student A.~V.~Radyushkin he accomplished the
factorization theorems for the meson form factors, linking
diagrammatic techniques with the operator product expansion (OPE).
The grounds for these works were supplied by previous investigations
by Efremov \cite{Efr74yad} and Efremov and collaborators
\cite{EG74fph}.

Moreover, Efremov and Radyushkin have diagonalized the anomalous
dimensions matrix for meson operators (in leading order) in terms of
Gegenbauer polynomials and first obtained the asymptotic distribution
amplitude (DA)
$\varphi(x,\mu^2\to\infty)\to\varphi^{as}(x)
=6x(1-x)$~\cite{ER80,LB79}.

Factorization theorems \cite{ER80,CZ77,LB79} make it possible to
calculate various hard processes in QCD involving mesons, in which the
meson DAs enter as the central nonperturbative input.

In the context of the present occasion, we are primarily interested in
presenting recent achievements in describing the pion characteristics
by mapping out its internal quark structure.
A short note on the nucleon is also included.

\section{Non-local condensates and pion distribution amplitude}
The pion DA of twist-2, $\varphi_\pi(x,\mu^2)$, is a gauge- and
process-independent characteristic of the pion that universally
specifies the longitudinal momentum $xP$ distribution of valence quarks
in the pion with momentum $P$
\begin{equation}
 \va{0\mid\bar{d}(0)\gamma^\mu\gamma_5E(0,z)u(z)\mid\pi(P)}
  \Big|_{z^2=0}
  = i f_{\pi}P^{\mu}
   \int^1_0 dx e^{ix(zP)}\
    \varphi_{\pi}(x,\mu^2)
\end{equation}
and where
$E(0,z)
=
{\cal P}\exp\!\left[-ig_s\!\!\int_0^z t^{a} A_\mu^{a}(y)dy^\mu\right]$
is a phase factor, path-ordered along the straight line connecting the
points $0$ and $z$ to preserve gauge invariance.

\subsection{Average QCD vacuum quark virtuality $\mathbf{\lambda_q^2}$}
The pion DA encapsulates the long-distance effects and therefore
reflects the nonperturbative features of the QCD vacuum.
The latter can be effectively parameterized in terms of non-local
condensates, as developed in \cite{MR86,BR91,MS93} by
A.~Radyushkin and two of us (A.B. and S.M.).
This provides a reliable method to construct hadron DAs that
inherently accounts for the fact that quarks and gluons can flow through
the QCD vacuum with \textit{non-zero} momentum $k_q$.
This means, in particular, that the \textit{average} virtuality of
vacuum quarks,
$\langle k_{q}^{2} \rangle = \lambda_{q}^{2}$
is not zero, like in the local sum-rule approach~\cite{CZ84}, but can
have values in the range~\cite{BI82}
$\lambda_{q}^{2} =
 \va{\bar{q}\left(i g\,\sigma_{\mu \nu}G^{\mu \nu}\right)q}
 /(2\va{\bar{q}q})=0.35-0.55~\gev{2}$.
Therefore, the non-local condensates in the coordinate representation,
say, $\langle \bar{q}(0)E(0,z)q(z)\rangle$,
are no longer \textit{constants}, but depend on the interval $z^2$
in Euclidean space and decay with the correlation length
$\Lambda \sim 1/\lambda_q$.
Lacking an exact knowledge of non-local condensates of higher
dimensionality, one has de facto to resort to specific Anz\"atze
\cite{BM02}, in order to parameterize the non-local condensates.
Nevertheless, it is important to stress that we were able to determine
in \cite{BMS02} $\lambda_q^2$ directly from the CLEO data \cite{CLEO98}
within the range predicted by QCD sum rules \cite{BI82} and
lattice simulations \cite{BM02}, favoring the value
$\lambda_{q}^{2} \simeq 0.4~\gev{2}$.

\subsection{QCD sum rules}
The distribution amplitudes $\varphi_{\pi (A_1)}(x,\mu^2)$ for the pion
and its first resonance can be related to the non-local condensates by
means of the following sum rule that is based on the correlator of two
axial currents
\begin{eqnarray}
\label{eq:nlcsrda}
  f_{\pi}^2\varphi_\pi(x) +
 f_{A_1}^2\varphi_{A_1}(x)\exp\{-\frac{m^2_{A_1}}{M^2}\}
&=&
  \int_{0}^{s_{\pi}^0}\rho^\text{pert}(x;s)e^{-s/M^2}ds +
  \frac{\langle \alpha_s GG\rangle}{24\pi M^2}\
   \Phi_G\left(x;M^2\right) \nonumber \\
&+&  \frac{8\pi\alpha_s\langle{\bar{q}q\rangle}^2}{81M^4}
      \sum_{i=S,V,T_1,T_2,T_3}\Phi_i\left(x;M^2\right)
\; ,
\end{eqnarray}
where the index $i$ runs over scalar, vector, and tensor condensates
\cite{BMS01,BM98}, $M^2$ is the Borel parameter, and $s_{\pi}^0$ is the
duality interval in the axial channel.
Above, the dependence on the non-locality parameter enters on the RHS
in the way  exemplified by the numerically important scalar-condensate
contribution
\begin{eqnarray}
\label{PhiS}
 \Phi_S\left(x;M^2\right)
  &=& \frac{18}{\bar\Delta\Delta^2}
       \Bigl\{
        \theta\left(\bar x>\Delta>x\right)
         \bar x\left[x+(\Delta-x)\ln\left(\bar x\right)\right]
       +  \left(\bar x\rightarrow x\right) + \nonumber \\
&&\qquad\quad
       + \theta(1>\Delta)\theta\left(\Delta>x>\bar\Delta\right)
         \left[\bar\Delta
              +\left(\Delta-2\bar xx\right)\ln(\Delta)\right]
         \Bigr\}
\end{eqnarray}
with $\Delta \equiv\lambda_q^2/(2M^2)$,
$\bar\Delta\equiv 1-\Delta$
and $\bar{x}\equiv 1-x$.
In the so-called \textit{local} approach~\cite{CZ84}, the end-point
contributions ($x\to 0$ or $1$) are strongly enhanced by
$\delta(x), \delta'(x) \ldots$ because they disregard the
finiteness of the vacuum correlation length $\Lambda$ 
by setting  in Eq.\ (\ref{PhiS})
$\lambda_q^2 \to 0$ to obtain
\begin{equation}
\label{limitS}
\lim\limits_{\Delta \to 0}\Phi_S\left(x;M^2\right)=
9\left[ \delta(x)+  \delta(1-x)\right].
\end{equation}
In contrast, taking into account the non-locality of the condensates
via $\lambda_q^2$, leads to a strong suppression of these regions.
Due to the end-point suppression property, the sum rule
(\ref{eq:nlcsrda}) allows us to determine the first ten moments
$\langle\xi^N\rangle_\pi \equiv \int_{0}^{1}
\varphi_\pi(x)(2x-1)^N dx$ of the pion DA
and \textit{independently} also the inverse moment
$\langle x^{-1}\rangle_{\pi}
\equiv \int_{0}^{1}\varphi_\pi(x)\ x^{-1} dx $
quite accurately (see in \cite{BMS01c,BMS03} for more details).
The intrinsic accuracy of this procedure admits to obtain the pion DA
moments with uncertainties varying in the range of 10\%.

\subsection{Models for the pion distribution amplitude}
Models for the pion DA, in correspondence to the extracted moments, can
be constructed
\begin{wrapfigure}[18]{l}{9cm}
 \epsfig{figure=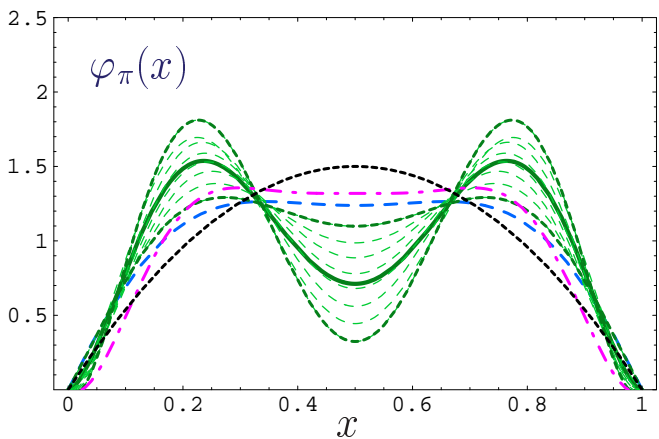,width=8.9cm}\\
\vspace{-8mm}
 \caption{\footnotesize Comparison of selected pion DAs denoted by
   obvious acronyms:
   $\varphi_\text{as}$ (dotted line) \protect\cite{ER80,LB79},
   $\varphi_\text{PR}$ (dashed line) \protect\cite{PR01},
   $\varphi_\text{Dor}$ (dot-dashed line) \protect\cite{Dor02}, and
   $\varphi_\text{BMS}$ (solid line) \protect\cite{BMS01}.
   Also shown is the whole ``bunch'' determined via QCD sum
   rules with non-local condensates \protect\cite{BMS01}.
   All DAs are normalized at the same scale
   $\mu_0^2\approx 1$ GeV$^2$.\vspace*{-1mm}}
\label{fig:pion_das}
\end{wrapfigure}
in different ways \cite{MR86,BM98}.
However, on the grounds explained above, it appears that two-parameter
models, the parameters being the first Gegenbauer coefficients
$a_{2}$ and $a_{4}$,
enable one to fit all the moment constraints for
$\langle \xi^{N} \rangle_\pi$, as well as to reproduce the value of
$\va{x^{-1}}_\pi$
within the QCD sum-rule error range, resulting into a ``bunch'' of DAs
displayed in Fig.\ \ref{fig:pion_das}.
The optimum sample out of this ``bunch'', termed BMS model,
is described by the following expression
\begin{eqnarray}
 \varphi^\text{BMS}(x)
  &=& \varphi^\text{as}(x)
       \left[1 + a_2\ C^{3/2}_2(2x-1)\right.
       \nonumber \\
              &&\ +  \left. a_4\ C^{3/2}_4(2x-1) \right]
 \label{optG}
\end{eqnarray}
with $a_2=+0.20$, $a_4=-0.14$
and is emphasized by a solid line 
in Fig.\ \ref{fig:pion_das}.
The shape of this ``bunch'' is confirmed 
by a non-diagonal correlator, 
based on the QCD sum rules considered in \cite{BM95}.

\section{CLEO data analysis}
The CLEO data \cite{CLEO98} on $F_{\pi\gamma}$ provide one rigorous
constraint on theoretical models for the pion DA in QCD.
Indeed, it was first shown in \cite{KR96} that these data exclude the
CZ pion DA because the prediction derived from it overshoots these data
by orders of magnitude.
Very recently, we analyzed \cite{BMS02,BMS03} the CLEO data by
combining attributes from QCD light-cone sum rules \cite{Kho99,SY99},
NLO Efremov--Radyushkin--Brodsky--Lepage (ERBL)
\cite{ER80,ER80tmf,LB79} evolution \cite{MR86ev,Mul94}, and
detailed estimates of uncertainties owing to higher-twist contributions
and NNLO perturbative corrections \cite{MMP02}.

The upshot of this analysis is that the CZ pion DA is excluded at the
4$\sigma$ level of accuracy 
and---perhaps somewhat surprisingly---that also the asymptotic pion DA 
\begin{wrapfigure}[25]{l}{9cm}
 \epsfig{figure=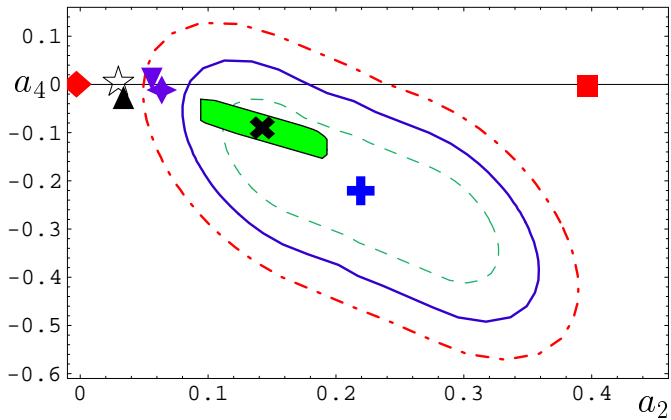,width=8.9cm}\\
\vspace{-8mm}
 \caption{\footnotesize Analysis of the CLEO data on
  $F_{\pi\gamma^{*}\gamma}(Q^2)$ in the ($a_2$, $a_4$) plane in terms
  of error regions around the best-fit point (blue cross) with the
  following designations:
  $1\sigma$ (broken green line);
  $2\sigma$ (solid blue line);
  $3\sigma$ (dashed-dotted red line).
  Various theoretical models are also shown for comparison.
  The designations are as follows: {\ding{117}}--the asymptotic DA,
   {\ding{54}}--BMS model, {\footnotesize\ding{110}}--{CZ} DA,
   {\ding{58}}--best-fit point, {\ding{73} \protect{\cite{PPRWK99}}},
   {\ding{70}~\protect{\cite{PR01}}}, {\small\ding{115}}
   \cite{ADT00}--instanton models, and
   {\footnotesize\ding{116}}--transverse lattice result
   \protect{\cite{Dal02}}.
  The slanted green rectangle represents the BMS ``bunch''
  of pion DAs dictated by the nonlocal QCD sum rules
  for the value $\lambda^2_q=0.4$~GeV$^{2}$.
  All constraints are evaluated at $\mu^2=5.76$~GeV$^2$
  after NLO ERBL evolution.
  }
\label{fig:ellipse}
\end{wrapfigure}
lies outside the 3$\sigma$ error ellipse
in the ($a_2,~a_4$) plane (see Fig.\ \ref{fig:ellipse}), even if one
allows the theoretical uncertainties owing to unknown higher-twist
contributions to be of the order of 30\% and presumes that the size of
NNLO perturbative corrections is also large.
On the other hand, the BMS pion DA calculated with a vacuum virtuality
$\lambda_{q}^{2} \simeq 0.4~\gev{2}$ was found to be inside the
1$\sigma$ error ellipse, while other rival models, based on differing
instantons approaches \cite{PR01,ADT00}, or derived with the aid of
lattice simulations \cite{Dal02}, 
are located in the vicinity of the
border of the 3$\sigma$ contour.
It is worth emphasizing that the more precise the instanton-based
models become, the further away from the asymptotic pion DA towards the
region of the ``bunch'' they move (we refer to \cite{BMS03} for more
details).\footnote{The new model relative to \cite{ADT00}, proposed in
\cite{Dor02}, involves more than two Gegenbauer coefficients and can
therefore not be displayed in Fig.\ \ref{fig:ellipse}. However,
reverting this model to an approximate one by utilizing  only two
(effective) Gegenbauer coefficients $a_2$ and $a_4$ shows that it is
close to the 3$\sigma$ error ellipse boundary, as said above.}
It was pointed out before in \cite{SSK99} that the CLEO data ask
for a broader pion DA than the asymptotic one.

\begin{figure}[t]
 \centerline{\includegraphics[width=0.48\textwidth]{%
   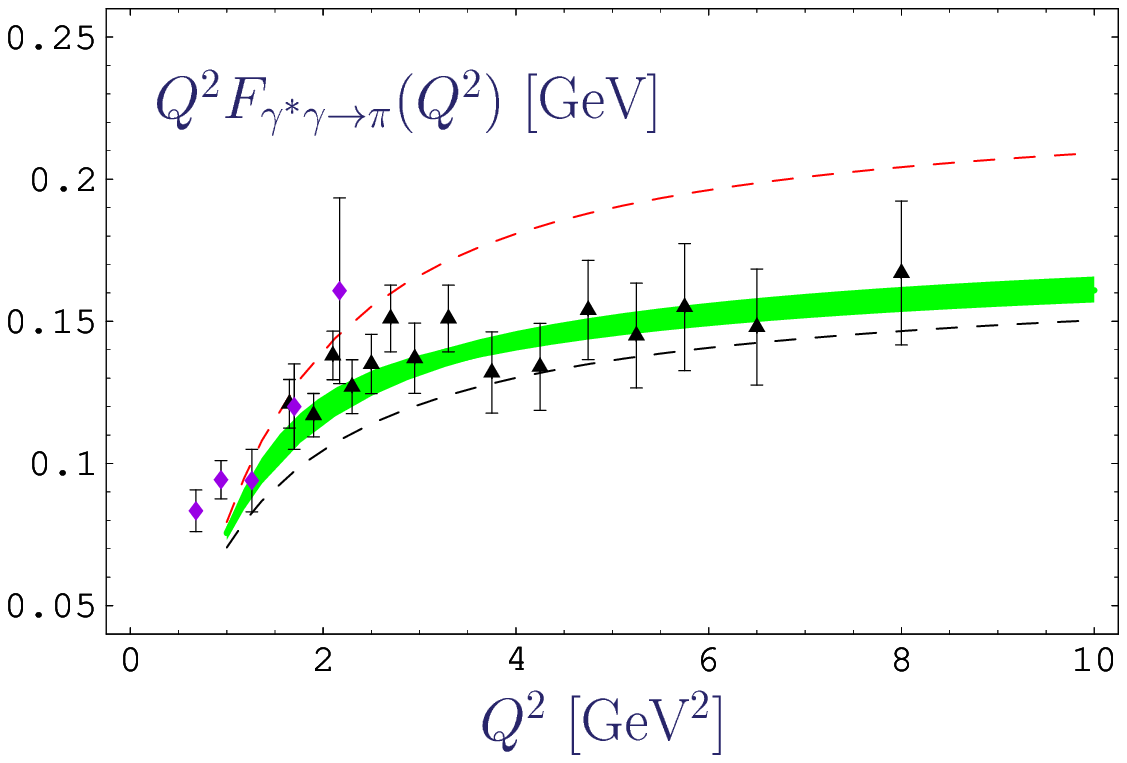}~%
  \includegraphics[width=0.48\textwidth]{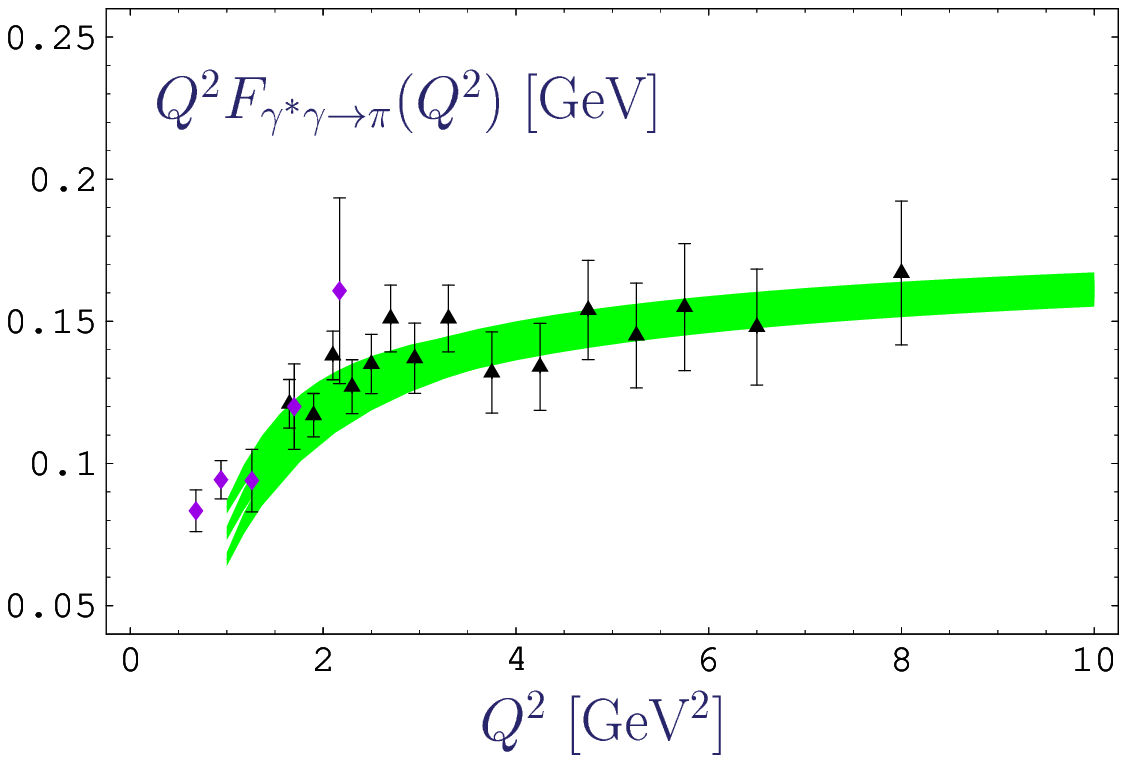}}
  \caption{\footnotesize
   Left: Light-cone sum-rule predictions for
   $Q^2F_{\gamma^*\gamma\to\pi}(Q^2)$
   in comparison with the CELLO (diamonds, \protect\cite{CELLO91})
   and the CLEO (triangles, \protect\cite{CLEO98}) experimental data
   evaluated with the twist-4 parameter value
   $\delta_\text{Tw-4}^2=0.19$~GeV$^2$~\protect\cite{BMS02,BMS03}.
   The predictions correspond to selected pion DAs; viz.,
   $\varphi_\text{CZ}$ (upper dashed line) \protect\cite{CZ84},
   BMS-``bunch'' (shaded strip) \protect\cite{BMS01}, and
   $\varphi_\text{as}$ (lower dashed line) \protect\cite{ER80,LB79}.
   Right: Our prediction for $Q^2F_{\gamma^*\gamma\to\pi}(Q^2)$
   corresponding to the ``bunch'' of pion DAs in
   Fig.\ \ref{fig:pion_das} (shaded strip) in comparison with
   experimental data for twist-4 parameter values varied in the range
   $\delta_\text{Tw-4}^2=0.15-0.23$~GeV$^2$.}
 \label{fig:Formfactor}
\end{figure}

In Fig.\ \ref{fig:Formfactor} (left panel) we compare our prediction
for the scaled pion-photon transition form factor with those from the
CZ model (upper dashed line) and the asymptotic DA (lower dashed line).
One observes that the strip obtained from the ``bunch'' of DAs is in
very good agreement with both the CLEO data and also with the CELLO
data \cite{CELLO91}.
The right panel of this figure illustrates in the form of a shaded band
the region of uncertainty induced by our limited knowledge of
higher-twist contributions.
One observes that even the low-$Q^2$ CELLO data are in reasonable
compliance with the theoretical prediction (the shaded strip).

Let us close this section by mentioning that other approaches claim to
be able to describe the CLEO data with the asymptotic pion DA
\cite{KR96,BJPR98,DKV01}, taking into account only the leading-twist
contribution and using only perturbative QCD (see for more details
in~\cite{BMS02}).

\section{Other exclusive processes}
\begin{figure}[t]
 \centerline{\includegraphics[width=0.48\textwidth,height=62mm]{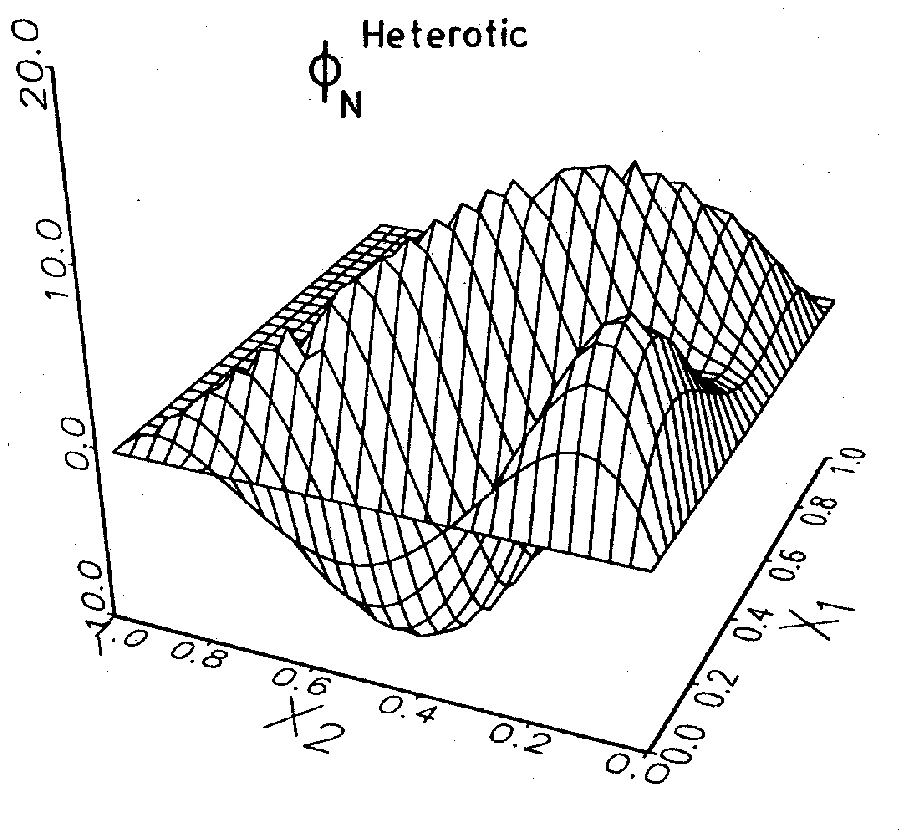}%
          ~~~\includegraphics[width=0.48\textwidth]{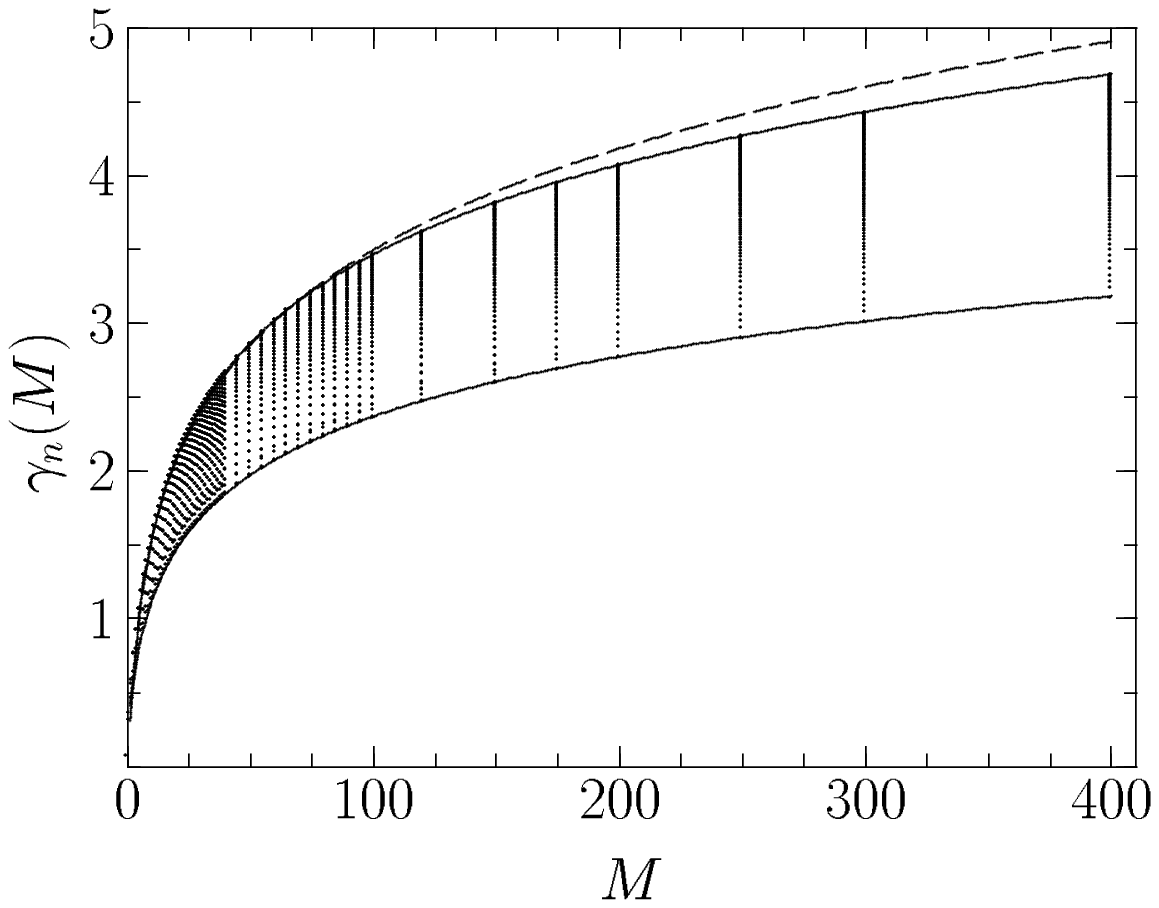}}
  \caption{\footnotesize Left: The heterotic nucleon distribution
         amplitude, proposed in \cite{SB93N}.
         Right: Spectrum of the anomalous dimensions of trilinear
         twist-3 quark operators up to order $M=400$.
         The solid lines (upper and lower envelopes of the spectrum)
         represent logarithmic fits up to the maximum considered order
         400, taking into consideration all orders above 10.
         The dashed line gives for comparison a previous logarithmic
         fit \cite{Ste99} which takes into account all orders up to
         150.
 \label{fig:spec400}}
\end{figure}

Factorization theorems can be extended---at least formally---to baryons
and their form factors \cite{LB79}.
The primary goal below is to give a brief summary of main results
rather than to review the subject and the status of individual
exclusive processes or baryon DAs (for a recent review, we refer to
\cite{Ste99}).
For instance, the situation concerning the nucleon DA is more
controversial compared to the meson case.
It is undoubtedly true that the asymptotic nucleon DA is unable to
describe the nucleon form factors \cite{LB79}.
On the other hand, asymmetric DAs constructed via moments determined by
\textit{local} QCD sum rules following \cite{CZ84}, as, for example, in
\cite{GS86,KS87,COZ89,Ste89,SB93N,BS93}
(see on the left part of Fig.\ \ref{fig:spec400} for an illustration),
seem to yield to strongly suppressed results for the magnetic nucleon
form factor when transverse momentum---intrinsic and Sudakov---effects
are included \cite{Ste95,Ste99}.
Valuable information on the inner structure of the nucleon was recently
provided in \cite{PP02} in the context of instantons, where it was
shown that the shape of the proton DA is far from the asymptotic one.

While the nonperturbative nature of the nucleon is yet not
well-understood, its evolution on the basis of the
renormalization-group equation can be performed to a high level of
accuracy within QCD perturbation theory.
Indeed, within the basis of symmetrized Appell polynomials
\cite{Ste94,Ste99}, the nucleon evolution equation can be solved by
employing factorization of the dependence on the longitudinal momentum
from that on the external (large) momentum scale $Q^{2}$ up to any
desired polynomial order.\footnote{The eigenfunctions of the nucleon
evolution equation are linear combinations of symmetrized Appell
polynomials, appropriately orthonormalized \cite{Ste99}.}
The spectrum of the corresponding anomalous dimensions of trilinear
quark operators was also determined \cite{Ste94,Ste99,BSS99} and its
large-order behavior seems to increase logarithmically, reflecting
the enhanced emission of soft gluons that forces the probability for
finding bare quarks to decrease (see Fig.\ \ref{fig:spec400}, right panel).
This spectrum can be reproduced by the logarithmic fit
\begin{equation}
 \gamma _{n}(M)
   = c + d \ln (M + b) \; .
 \label{eq:gammalog}
\end{equation}
The upper envelope of the spectrum is best described by the following
values of the parameters with their errors:
$b=1.90989 \pm 0.00676$,
$c=-0.637947 \pm 0.000634$, and
$d = 0.88822 \pm 0.000119$.
For the lower envelope, the corresponding values are
$b=3.006 \pm 0.483$,
$c=-0.3954 \pm 0.0290$, and
$d = 0.59691 \pm 0.00545$.
The spacing of eigenvalues at very large order is reproduced by the
values $b=-0.027 \pm 0.728$,
$c=-0.2460 \pm 0.0248$, and
$d = 0.291883 \pm 0.00475$.
For every order $M$, there are $M+1$ eigenfunctions of the same order
with an excess of symmetric (under the permutation $P_{13}$) terms
(denoted by black dots in Fig.\ \ref{fig:spec400}) by one for even
orders.
The total number of eigenfunctions up to order $M$ is
$n_{\rm max}(M)=\frac{1}{2}(M+1)(M+2)$ and the corresponding $(M+1)$
eigenvalues are obtained by diagonalizing the $(M+1)\times (M+1)$
matrix.
Up to order 150, both sectors 
(corresponding to the permutation parity $S_{n}=\pm 1$) 
of eigenvalues are included.
Beyond that order, for reasons of technical convenience, only the
antisymmetric (open circles) ones have been taken into account.
The multiplet structure of the anomalous dimensions spectrum was
found independently later on \cite{BDKM99} in the context of a
Hamiltonian approach to the one-dimensional XXX Heisenberg spin
magnet of non-compact spin $s=-1$.

\section{Concluding remarks}
Our discussion of the pion DA in the context of QCD sum rules with
non-local condensates shows that the vacuum non-locality parameter
can serve to extract valuable information on the underlying
nonperturbative dynamics.
The double-humped shape with suppressed end-points of the derived
pion DA is in good agreement with the CLEO data with a 1$\sigma$
accuracy and agrees with the CELLO data as well.
Progress of the non-local sum-rules approach to encompass tree-quarks
states, like the nucleon, appears promising, while the perturbative
apparatus for the evolution of such DAs is already well-developed.

In conclusion, let us mention as a personal statement that the major
part of our scientific work depends to a great extent on the power of
factorization theorems and their usage in QCD in the context of form
factors, structure functions, etc.
Therefore, we feel particularly attached to Prof.~Efremov, given also
that he was the Leader of the BLTPh QCD group, where two of us 
(A.P.B. and S.V.M.) have been working for over a decade, 
and he was also one of the opponents of one of us (N.G.S.) 
in defending his \textit{Doctor fiziko-matematicheskih nauk} degree.

\bigskip

\noindent \textbf{Acknowledgments.}
One of us (N.G.S.) is indebted to members of BLTPh, JINR for the warm
hospitality extended to him during a stay when this contribution was
prepared.
This work was supported in part by the Deutsche Forschungsgemeinschaft
(436 RUS 113/752/0-1), the Heisenberg--Landau Programme (grant 2003),
the Russian Foundation for Fundamental Research
(grants No.\ 03-02-16816 and 03-02-04022),
and the INTAS-CALL 2000 N 587.



\end{document}